\documentclass [twocolumn,showpacs,amssymb,prd]{revtex4}

\usepackage{graphicx}
\usepackage{dcolumn}
\usepackage{bm}
\usepackage{epsfig}

\begin{document}
\newcommand{\gsim}{\hbox{\rlap{$^>$}$_\sim$}}
\newcommand{\lsim}{\hbox{\rlap{$^<$}$_\sim$}}

\title{On The Missing Counterparts \\
Of LIGO-Virgo Binary Merger Events} 
\author{Shlomo Dado and Arnon Dar}
\affiliation{Physics Department, Technion, Haifa 32000, Israel}

\begin{abstract}
Despite world-wide ground, underground, and space based observations 
in search of anticipated electromagnetic and perhaps neutrino 
counterparts to the 29 compact binary merger events, which have been 
detected by the upgraded LIGO-Virgo gravitational wave detectors  
in the first half year of their observation period O3, no such 
counterparts were found. Although the current situation could be due to 
a poor localization of nearby merger events and/or a 
complex background of short extragalactic transients, it could also 
be intrinsic. We show that, indeed, it is quite expected in the cannonball 
model of gamma ray bursts and their afterglows. 
\end{abstract}

\maketitle
The short-duration gamma ray burst (SGRB) 170817A [1] that followed 
$1.74\!\pm\! 0.05$ s after the end of the gravitational wave (GW) emission 
event GW170817, which was detected by LIGO-Virgo [2], has shown beyond 
doubt that neutron star mergers produce SGRBs [3] with an afterglow (AG) that 
extends over the radio [4], optical [5], and x-ray bands [6].  However, 
despite the world-wide campaigns to detect the anticipated electromagnetic 
counterparts, and perhaps a neutrino counterpart, to the 29 binary merger 
events, which were detected by the upgraded LIGO-Virgo gravitational wave 
detectors [7] in the first half year since the beginning of their third 
observation period (O3)  -4 neutron stars binaries (NSBs), 4 neutron star 
- black hole binaries (NSBHs), and 21 binary black holes (BBHs)- no such 
counterparts have been detected. In this letter we argue that such a 
situation is that expected in the cannonball (CB) model of GRBs [8] if 
most of the NSB mergers, like all the NSBH and BBH mergers, produce a 
black hole remnant rather than a neutron star remnant.

In the CB model of SGRBs [9], mass accretion episodes of fall back matter 
on the newly born compact object  -a neutron star, a quark star, or a 
black hole in NSB merger events, or a black hole in NSBH and BBH events-  
produce narrowly collimated bipolar jets of plasmoids (cannonballs) with 
a bulk motion Lorentz factor $\gamma\!\gg\!1 $ and a Doppler factor 
$\delta\!=\!1/\gamma(1\!-\!\beta\, cos\theta)$, where $\theta$ is the 
viewing angle of the jet  relative to the jet direction of motion. 
These highly relativistic CBs produce 
narrowly collimated beams of gamma rays by inverse Compton scattering 
(ICS) of photons surrounding the newly born compact object.
This simple model has been very successful in predicting  the main 
observed properties of the prompt gamma ray emission in  long
and short duration  GRBs.  They provide compelling evidence that 
the prompt gamma ray emission in GRBs is narrowly beamed, and 
consequently, most of the GRBs do not point in the direction of Earth.  

Perhaps the best indirect evidence that GRBs are narrowly beamed 
is provided by the observed correlations between their main properties.
For instance, in the CB model the produced beams of gamma rays  have 
an opening angle $\theta\!\approx\!1/\gamma\ll 1$, an observed peak energy 
$E_p\!\approx\!\epsilon_p \gamma \delta/(1\!+\!z)$ and an isotropic 
equivalent energy $E_{iso}\!\propto\!\epsilon_p \gamma\delta^3$ in the 
burst rest frame, where $\epsilon_p$ is the peak energy of the photons 
surrounding the newly born compact object. For $\gamma\!\gg\! 1$ and 
$\theta^2\!\ll\!1$, $\delta\!\approx\!2\,\gamma/(1\!+\!\gamma^2\theta^2)$, 
and consequently the most probable viewing angles of distant GRBs is 
$\theta\!\approx\!1/\gamma$, which yield $\delta\!\approx\! \gamma$, 
and, consequently, the [$E_p,E_{iso}$] correlation [10] in ordinary GRBs,
\begin{equation}
(1+z)\,E_p\propto [E_{iso}]^{1/2},
\end{equation}
which was discovered empirically [11]. 

Moreover, in the CB model, $E_{iso}$ of GRBs, which are viewed from far off-axis 
(FOA), i.e., from angles which satisfy $\gamma^2\theta^2\!\gg\!1 $, depends 
strongly on the suppressed  value $\delta\!\approx\! 2/\gamma \theta^2$. Such GRBs 
have much smaller  values of $E_{iso}$ and peak luminosity $L_p$ compared to 
GRBs which are viewed from near axis (NA):
\begin{equation}
E_{iso}({\rm FOA})/E_{iso}({\rm NA})\!\approx\! (\gamma^2\,\theta^2/2)^{-3}\!\ll\! 1, 
\end{equation}
and since the observer time $t$ and the time $t'$ in the GRB rest frame are related 
by $dt\!=\!(1+z)dt'/\gamma\delta$,  
\begin{equation}
L_p({\rm FOA})/L_p({\rm NA}) \approx (\gamma^2\,\theta^2/2)^{-4}\!\ll\! 1.
\end{equation}  
Consequently, in the CB model far-off axis GRBs are low luminosity GRBs,
which satisfy the correlation (eq.(35)in [10])
\begin{equation}
(1+z)\,E_p\propto [E_{iso}]^{1/3}.
\end{equation}
Figure 1 demonstrates that bright SGRBs and low-luminosity 
SGRBs, satisfy the same  $[E_p,E_{iso}]$ 
correlations, which are satisfied by
bright and low luminosity GRBs, and 
are given, respectively, by eqs.(1) and (4),
as predicted by the CB model of 
GRBs [8,9]. Many more tests of the collimated nature of GRBs 
are reviewed in [12].   

\begin{figure}[]
\centering
\epsfig{file=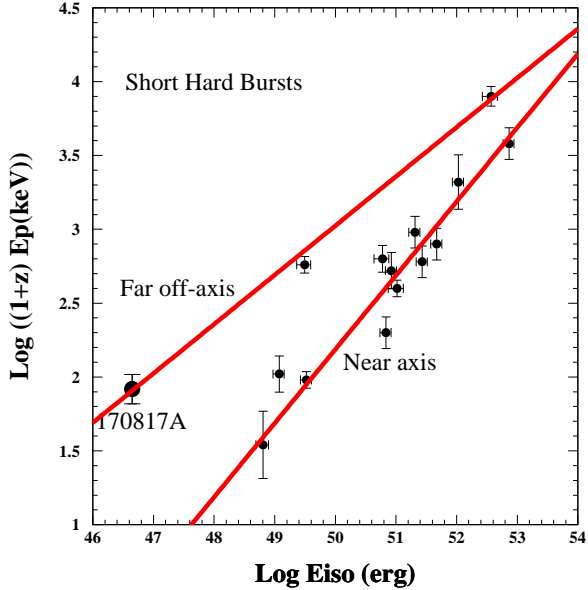,width=8.6cm,height=8.6cm}
\caption{The $[E_p,E_{iso}]$ correlations in SHBs.
The lines are the CB model predicted correlations, as 
given by eqs.(1),(4).}
\end{figure}

GRBs are detectable only if their energy flux is above the 
detection threshold of the space based GRB detectors.
Because of beaming, the observed peak energy flux 
of far off-axis GRBs, decreases rapidly, like $L_p\!\propto \theta^{-8}$,  with 
increasing viewing angle $\theta$, as given by eq.(3). 
Consequently, only a small fraction 
of the SGRBs which are produced by merger events are detectable. 
This fraction decreases rapidly with increasing distance. 
GRB170817A, however, which was 
detected [1] 1.74 s after the NSB merger 
event GW170817 [2], was a rare  SGRB which pointed 
in a direction far away from  Earth ($\theta\!\approx\!28^{\circ}$ [13]).
It was detected at such large viewing angle 
because it took place in a relatively very nearby galaxy ($z\!=\!0.0096$ [14])
and as expected (see Figure 1) had an  extremely low values of  $E_{iso}$ and $L_p$.

In the CB model, the  above considerations, which apply to  merger events which produce SGRBs,
also apply to their narrowly beamed counterparts which include  their  extended 
emission (EE) (see Figure 2) and their beamed afterglows. 
\begin{figure}[]
\centering
\vspace{1.cm}
\epsfig{file=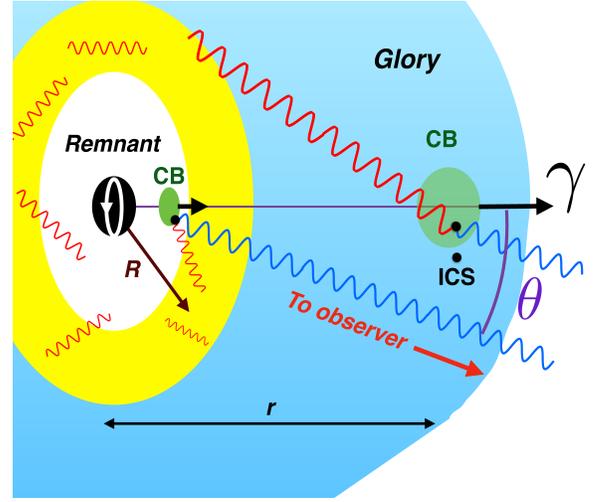,width=7.6cm,height=6.6cm}
\caption{A CB as it crosses the light
(glory) surrounding the merger remnant. The CB's
electrons Compton up-scatter glory photons
with incident angles which decrease with 
increasing distance from the CB launch site 
within the glory. ICS of glory photons 
within the tidally disrupted neutron star matter 
(yellow torus) produces a prompt emission pulse, while ICS  
of photons of an extended  nebula or a star cluster
produces the extended emission.}
\label{Fig2}
\end{figure}

However, the afterglow of SGRBs seems to 
include also  an early time isotropic component -  
nebular  emission powered by the spin down of 
a neutron star remnant of the NSB merger.
This  is supported  by the  successful reproduction [15] 
of the  X-ray light curves of SGRBs with the extended emission  followed by 
a well sampled early time X-ray afterglow, as demonstrated below.

In the CB model, such light curves of the observed extended emission 
plus the early time X-ray afterglow powered by a millisecond pulsar (MSP), 
above a minimum photon 
energy $E_m$, have the approximate X-ray lightcurve,
\begin{equation}
\int_{E_m}E{d^2N(E,t)\over dE\,dt}dE\!\approx\!
 {A_{ee}\,exp(\!-\!\beta\!\sqrt{1\!+\!(t/\tau_{ee})^2})\over  
 1\!+\!(t/\tau_{ee})^2}+
               {A_{msp}\over (1\!+\!t/t_b)^2}
\end{equation}
where $t$ is the observer time since the beginning of the burst. In eq.(5),
the first term on the right hand side (RHS) is the beamed EE contribution 
due to ICS of photons in the extended glory around the merger site by the
highly relativistic bipolar jet. The second term on the RHS is the early 
time isotropic nebular afterglow powered by the spin down of the newly born MSP
[15]. In eq.(5), we have assumed, for simplicity, that the pulsar wind nebula (PWN) 
is a torus   
perpendicular to the highly relativistic bipolar jet, with a radius $R$ and
with the merger site at its center, as shown in Figure 2. We have also
assumed that the glory has approximately a bremmstrahlung spectrum, i.e.,
an exponentially cut off power law (CPL) spectrum,
$dn_\gamma/d\epsilon\!\propto\!\epsilon^{-\alpha}\,exp(-\epsilon/kT)$, at
redshift $z$, with $\alpha\!\approx\!1$ and temperature $T$.
The density of glory photons
decreases as function of distance $d$ from the merger site along the jet
trajectory, roughly as $n(d)\!=\!n(0)\,/(1\!+\!d^2/R^2)$ where
$d\!\approx\! c\,\gamma\,\delta\,t/(1\!+\!z)$ and $t/\tau_{ee}\!=\!d/R$.
At $d\!=\!0$, the glory photons incident at an angle 90$\deg$ on the CB,
which increases with increasing $d$ and yield an ICS
$\beta\!=\!(1\!+\!z)E_m/\gamma\delta\,kT$.

Figures 3,4 show best fits of eq.(5), based on the CB model of SGRBs, 
to the observed well sampled early time x-ray light curves measured 
with the Swift X-Ray Telescope (XRT) [16] for  two representative SGRBs, 
060614 and 150424A.  The best fit parameters for GRB060614,  
$\beta\!=\!0.48 $, $t_{ee}\!=\!78.9$ s,  $t_b\!=\!48918 $, 
and $A_{msp}/A_{ee}\!=\!0.40E-4$, yield $\chi^2/dof\!=\!679/509\!=\!1.33$, 
and the best fit parameters, $\beta\!=\!0.47 $, $t_{ee}\!=\!116.5$ s, 
and $t_b\!=\!28050$ s,  and $A_{msp}/A_{ee}\!=\! 1.494$  yield
$\chi^2/dof\!=\!186/127\!=\!1.46$ for GRB150424A. 
These fits and  those published in [15] strongly support 
the existence of an isotropic afterlow component in NSB mergers 
which produce MSP remnants.

\begin{figure}[]
\centering
\epsfig{file=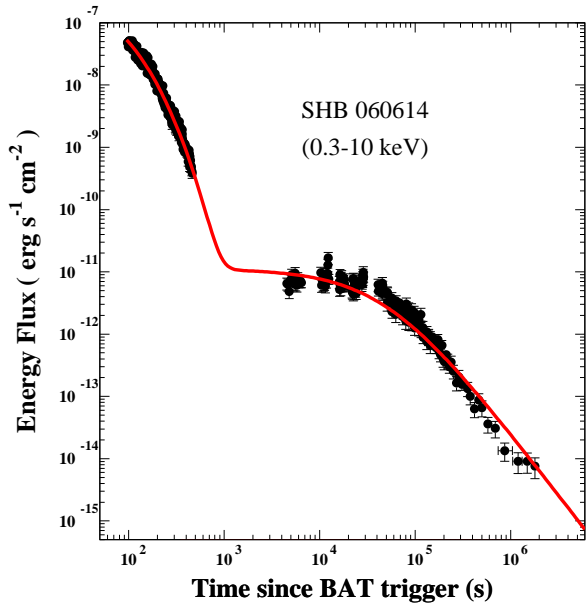,width=8.6cm,height=8.6cm}
\caption{The x-ray light curve of the extended 
emission plus the early time afterglow emission of SGRB060614
measured with the Swift x-ray telescope [16] and the  best fit 
light curve given by eq.(5) for a beamed extended emission 
followed by an isotropic nebular emission 
powered by a newly born MSP. The best fit has 
a  $\chi^2/dof\!=\!679/509\!=\!1.33$.}
\end{figure}

\begin{figure}[]
\centering
\epsfig{file=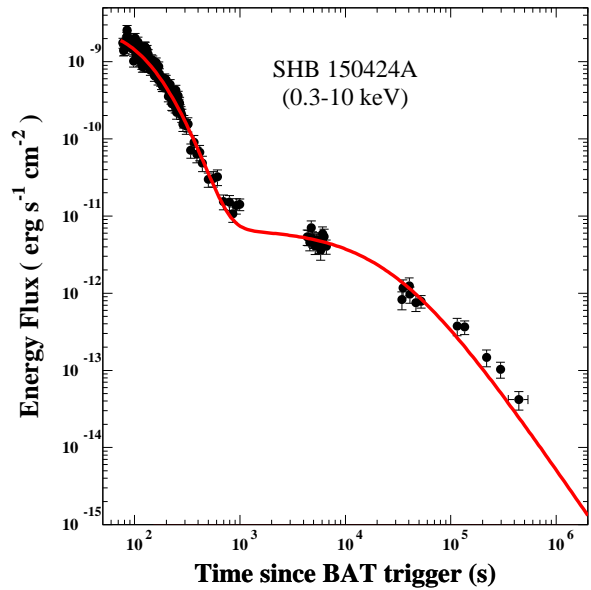, width=8.6cm,height=8.6cm}
\caption{The x-ray light curve of the extended
emission plus the early time afterglow emission of SGRB150424A
measured with the Swift x-ray telescope [16] and the  best fit
light curve given by eq.(5) for a beamed extended emission   
followed by isotropic nebular emission
powered by a newly born MSP. The best fit has 
$\chi^2/dof\!=\!186/127\!=\!1.46$.}
\end{figure}

An isotropic nebular emission component powered by the spin down 
of a newly born MSP is produced only in NSB merger events with a neutron star remnant.  
NSB merger events with a black hole remnant produce only narrowly beamed 
electromagnetic counterparts, most of which are beamed far off axis. Such 
events are much more frequent among the LIGO-Virgo NSB events than those which 
produce a neutron star remnant that powers an isotropic nebular emission. This 
is because the gravitational wave luminosity depends on the masses of the 
merging neutron stars, which is given roughly by 
\begin{equation} 
L(GW)\!\approx\! {32\over 5}{G^4\,\mu^2 M^3\over c^5r^5},
\end{equation} 
where $M\!=\!M1\!+\!M2$ is their total mass, $\mu\!=\!M1\!\times\!M2/M $ is 
their reduced mass, and $r$ is their separation. The production of a black hole 
remnant requires larger 
masses of the merging neutron stars than those in the production of a neutron 
star remnant. Consequently, the LIGO-Virgo detection range of NSB merger 
events which produce black hole remnants, is larger than the detection range 
of NSB mergers which produce MSP remnants. Thus, the NSB merger events 
detected by LIGO-Virgo, probably are strongly enriched with NSB merger events 
with a remnant black hole. Such mergers produce only narrowly beamed SGRBs 
with narrowly beamed extended emission and afterglow, most of which do not 
point close enough to Earth, and thus avoid detection.\\
{\bf Conclusions:} 
By chance, the rare proximity [14] of the NSB which produced GW170817 and the 
pointing direction of its rotation axis, made still possible the detection of its 
beamed electromagnetic counterparts, SGRB170817A and its afterglows, as well as 
its isotropic nebular emission, despite being a very far off-axis SGRB.

Most of the electromagnetic and neutrino counterparts of merger events which 
produce a remnant black hole, probably, are narrowly beamed. Most of them do not 
point in the direction close enough to Earth. Millisecond pulsar remnants of NSB 
mergers power also a nebular isotropic emission. But, because of their relatively 
lower GW luminosity compared to that of NSB mergers which produce black hole 
remnant, their detection range and consequently detection rate by LIGO-Virgo, are 
strongly reduced compared to those of NSB mergers which produce a black hole 
remnant. This situation is summarized in Table I.

Beaming of most of the SGRBs, their extended emissions, and afterglows in 
directions which do not point to Earth, most probably is the main physical reason 
why these electromagnetic counterparts of LIGO-Virgo merger events are missing. 
Althogh there is mounting indirect evidence that GRBs and their afterglows are 
narrowly beamed [8], so far the only direct observational evidence for that was the 
very large base interferometric (VLBI) observations of an apparent compact 
superluminal radio source (CB ?) moving away from the site of the GW170817 event 
[17] in a direction consistent [18] with that of the rotational axis of the merging 
NSB which produced it [2]. If and when the directions of the rotation axis of 
compact binaries which produce LIGO-Virgo binary merger events will be extracted 
with sufficient accuracy from the GW observations it may be possible to test the 
validity of the beaming explanation (assuming that SGRBs are pointing along this 
axis, which was roughly verified [18]  in  the  GRB170817A/GW170817 event.)

All the above arguments are also valid if the assumed nebular emission 
in NSB mergers with a remnant MSP is replaced by the emission of 
a very bright kilonova [19]. 

{}

\begin{table*}
\caption{Electromagnetic counterparts (CP) of binary mergers}
\label{table1}
\centering
\begin{tabular}{l l l l l l l l l}
\hline
Binary &Remnant&~~GRB   &~~EE    & ~~AG   &Remnant's AG &Detectable CP\\ 
\hline
NSB    &~~~NS  & Beamed & Beamed & Beamed &~~~~Isotropic  &~~~~Frequent  \\
NSB    &~~~BH  & Beamed & Beamed & beamed &~~~~~~None     &~~~~~Seldom   \\     
NSBH   &~~~BH  & Beamed & Beamed & beamed &~~~~~~None     &~~~~~Seldom   \\
BBH    &~~~BH  & Beamed & Beamed & beamed &~~~~~~None     &~~~~~Seldom   \\
\hline
\end{tabular} 
\end{table*}
\end{document}